\documentclass[conference,10pt]{IEEEtran}
\usepackage{xcolor}
\usepackage{mathtools}
\usepackage{epsfig,makeidx,color}
\usepackage{amsmath,amssymb,bbm}
\usepackage{cite,graphicx,lipsum}
\usepackage{enumerate}
\usepackage[switch,pagewise]{lineno}
\usepackage{hyperref}
\hypersetup{
        colorlinks = true,
        citecolor=blue,
}
\def\cX{{\cal X}}

\def\uE{{\mathbb E}}

\DeclareMathOperator*{\argmin}{\arg\!\min}

\newtheorem{myexample}{\it Example} 

\def\be{ \begin{equation} }
\def\ee{ \end{equation} }
\def\bea{ \begin{eqnarray} }
\def\eea{ \end{eqnarray} }

\def\bx{{\bf x}}

\def\b0{{\bf 0}}

\def\cI{{\cal I}}

\def\sH{{\sf H}}
\def\sI{{\sf I}}

\ifCLASSOPTIONonecolumn
  \newcommand{\figwidth}{0.50\columnwidth}
  
\else
  \newcommand{\figwidth}{0.90\columnwidth}
  
\fi

\usepackage{geometry}
\begin{document}

\IEEEoverridecommandlockouts

\title{A Unified View on Semantic Information and Communication: A Probabilistic Logic Approach}
\author{
\IEEEauthorblockN{Jinho Choi, Seng W. Loke, and Jihong Park\thanks{This research was supported
by the Australian Government through the Australian Research
Council's Discovery Projects funding scheme (DP200100391).}}
\IEEEauthorblockA{School of Information Technology\\
Deakin University, Australia \\
Email: \{jinho.choi, seng.loke, jihong.park\}@deakin.edu.au}
}

\maketitle
\begin{abstract}
This article aims to provide a unified and technical approach to semantic information, communication, and their interplay through the lens of probabilistic logic. To this end, on top of the existing technical communication (TC) layer, we additionally introduce a semantic communication (SC) layer that exchanges logically meaningful clauses in knowledge bases. To make these SC and TC layers interact, we propose various measures based on the entropy of a clause in a knowledge base. These measures allow us to delineate various technical issues on SC such as a message selection problem for improving the knowledge at a receiver. Extending this, we showcase selected examples in which SC and TC layers interact with each other while taking into account constraints on physical channels.
\end{abstract}


\ifCLASSOPTIONonecolumn
\baselineskip 28pt
\fi

\section{Introduction}

Communication systems have been significantly transformed over the last decades, yet the foundation of the underlying information and communication technology has been consistently laid by Shannon's theory \cite{Shannon}. In his theory, information is characterized as randomness in variables. This allows one to calculate the fundamental limit and performance of communication, and to design efficient compression and transmission schemes through noisy channels. Despite the success in this domain of \emph{technical communication (TC)}, since its introduction in 1948, Shannon theory's ignorance about the meanings of information \cite{Weaver53} has long been tackled particularly in the field of the philosophy of information. Meanwhile, overcoming this limitation of Shannon theory has recently been regarded as one of the key enablers for the upcoming sixth generation (6G) communication systems \cite{Strinati21,dde,seo2021semantics}. 

To fill this void, it requires to develop a theory on meaningful information, i.e., \emph{semantic information}, as well as a novel communication technology based on semantic information, i.e., \emph{semantic communication (SC)}. For SC, existing works can be categorized into model-free methods leveraging machine learning \cite{dde}, and model-based approaches that quantify semantic information \cite{Bao11} or specify the emergence of meanings through communication \cite{seo2021semantics}. Our work falls into the latter category in the hope of unifying our analysis on SC with the existing model-based analysis on TC.

In regard to semantic information, there are two different views in the philosophy of information. One angle focuses on measuring semantic similarity \cite{Floridi05,Floridi08}, which often encourages an entirely new way to define meaningful information. For instance, each meaning can be identified as a group that is invariant to various nuisances (e.g., a so-called topos in category theory \cite{belfiore2021topos}), across which semantic similarity can be compared.
The other end of the spectrum focuses on quantifying semantic uncertainty \cite{Adriaans10}, in a similar way to Shannon theory where message occurrences are counted to measure semantic-agnostic uncertainty. As an example, Shannon information can be extended to semantic information by leveraging the theory of inductive probability \cite{Carnap50} (see also \cite{Hailperin84,Williamson02}). This allows to measure the likelihood of a sentence/clause's truth using logical probability, upon which an SC system can be constructed~\cite{Bao11}. Our view is aligned with the latter angle (i.e., like \cite{Bao11}, a probabilistic logic approach is taken), 
while we focus on making SC interact with TC under Shannon theory.





In particular, in this paper, 
we consider an approach to SC based on the theory of probabilistic logic assigning probabilities to logical clauses \cite{Carnap50,Nilsson86}. This allows to make inferences over clauses and to quantify their truthfulness or provability in a probabilistic way. We showcase that the process of inference and its provability analysis can be performed using ProbLog\footnote{ProbLog tools are available in: \url{https://dtai.cs.kuleuven.be/problog}.}, a practical logic-based probabilistic programming language that has been widely used in the field of symbolic artificial intelligence (AI). 

Furthermore, based on  \cite{Bar-Hillel53,Adriaans10}, we consider a two-layer SC system comprising: (i) the conventional TC layer where data symbols can be transmitted without taking into account their meanings; and (ii) an SC layer where one exploits semantic information that can be obtained from a background knowledge or by updating a knowledge base. We demonstrate the interaction between TC and SC layers with selected examples showing how SC improves the efficiency of TC, i.e., \emph{SC for TC}, as well as how to design TC to achieve maximal gains in SC under limited communication resources, i.e., \emph{TC for SC}. 
For simplicity and consistency throughout the paper, we confine ourselves to a simple scenario where a human user or an intelligent device stores logical clauses in a knowledge base and intends to improve the knowledge by seeking answers to a number of queries.



\section{Background}    \label{S:Background}

In this section, we briefly present a background on information theory \cite{CoverBook} and probabilistic logic \cite{Nilsson86}.

\subsection{Preliminaries for Information Theory}

Although information theory originally started as a mathematical theory for communications, it
has been applied in diverse fields ranging from biology to neuroscience.
In information theory, random variables are used to represent symbols to be transmitted. The entropy of a random variable, denoted by $X$, is the number of bits required to represent it, which is given by $\sH(X) = - \sum_x \Pr(x) \log \Pr(x) = \uE[ - \log \Pr(X)]$ (taking $\log$ to base 2 in the rest of the paper) when $X$ is a discrete random variables, where $\Pr(x)$ stands for the probability that $X = x$ and $\uE[\cdot]$ represents the statistical expectation. 
The entropy of $X$ can also be interpreted as the amount of information of $X$.

The joint entropy of $X$ and $Y$ is defined as $\sH(X, Y) = \uE[-\log \Pr(X,Y)]$ and the conditional entropy is given by
$$
\sH(X\,|\, Y) = \uE[ - \log \Pr(X\,|\, Y)] = \sH(X,Y) - \sH(Y).
$$
The mutual information between $X$ and $Y$ is defined as $\sI(X;Y) =
\uE\left[\log \frac{\Pr(X,Y)}{\Pr(X) \Pr(Y)} \right]$. It can also be shown that $\sI(X;Y) = \sI(Y;X) = \sH(X) - \sH(X\,|\, Y) = \sH(Y) - \sH(Y\,|\, X)$. 
If $X$ and $Y$ are assumed to be the transmitted and received signals over a noisy channel, $\sI(X;Y)$ can be seen as the number of bits that can be reliably transmitted over this channel. Thus, $\max_{\Pr(x)} \sI(X;Y)$ is called the channel capacity that is the maximum achievable transmission rate for a given channel that is characterized by the transition probability $\Pr(Y\,|\, X)$. 

As pointed out in \cite{Bar-Hillel53}, information theory is not interested in the content or meaning of the symbols, but quantifying the amount of information based on the frequency of their occurrence (i.e., the distribution of symbols as random variables). For example, $\sH(X)$ is to measure the amount of information or number of bits to represent a symbol $X$ regardless of what $X$ means. 
However, this does not mean that information theory is useless in dealing with the meaning or content of information as will be discussed in the paper.

\subsection{Preliminaries for Probability and Logic}

Following the theory of probabilistic logic, we assign probabilities to logical clauses, and carry out probabilistic reasoning using a practical logic programming language, ProbLog. In ProbLog, each logical clause (e.g., rules or facts) is annotated with a probability (by a programmer) that indicate the degree of (the programmer's) belief in the clause.

Precisely, for facts $a$ and $b$, where $a$ is assigned probability $p_a$ and $b$ is assigned probability $p_b$, we have probability of $a \wedge b$ computed as the product of the probabilities, i.e. $p_a \cdot p_b$, and $a \lor b$ computed as $1-(1-p_a)\cdot(1-p_b)$ since $a \lor b = \neg (\neg a \wedge \neg b)$. 
Similar calculations can be applied with deductive reasoning, e.g., suppose  we have the rule $r$ of the form $a \rightarrow b$ (where ``$\rightarrow$'' is ``implies'')  annotated with probability $p_r$ and $a$ with probability $p_a$, then we can infer $b$ with probability $p_r \cdot p_a$. In ProbLog, a clause $a \rightarrow b$ with probability $p$ is written as {\tt p::b~:-~a}, where ``{\tt :-}'' can be read as ``if''.

In general, given a knowledge base $K$ is regarded as a set of clauses (where a clause is a rule or a fact). Given a rule of the form $a \rightarrow b$, the head of the rule is $a$ and the body is $b$. Note that a fact is basically a rule of the form $a \rightarrow true$, which can just be written as $a$. 
One can make inferences about the truth value of a query $q$, provided that $q$ matches the head of a clause in $K$ with the outcome being  the probability of $q$. If $q$ does not match any head of a clause in $K$, $K$ cannot say anything about $q$. We denote the probability of $q$ computed as the answer when posed as a query to the knowledge base $K$ by $p[K \vdash q]$. We assume that inferences made will be as defined by the semantics of ProbLog.

In addition, for the purposes of the discussion in this paper, we consider mostly the propositional logic fragment of ProbLog for simplicity (and if variables are involved in some examples, we assume that their values range over a finite set, i.e., they are just abbreviations for a finite set of propositional clauses, so that the set of queries that can be answered via a knowledge base is finite).

\section{Entropy and Knowledge Bases: Communicating  Informative Messages} \label{S:EKB}

\subsection{Entropy of a Clause}
We  consider the entropy $\sH_f$ of a given clause  $c$  whose truth value can be regarded as a random variable with outcomes ``true'' with probability $p_c$, and ``false'' with probability $1-p_c$,   as follows:
\[
 \sH_f(c) = - \left(p_c  \log (p_c) + (1-p_c)  \log(1-p_c) \right).
\]
Here, the subscript $f$ is used to differentiate the entropy of a random variable from that of a clause.

When a given query $q$ is posed to the knowledge base $K$, and suppose a probability $p_q$ is computed with respect to   $K$, i.e., when $q$ matches a clause's head  in $K$, as in the semantics of ProbLog, then  $p_q = p[K \vdash q]$, and we denote the entropy of $q$ with respect to $K$ as   $\sH_f^K(q)$, i.e.:
\[
 \sH_f^K(q) = - \left(p_q  \log (p_q) + (1-p_q)  \log(1-p_q) \right).
\]
Note that if $q$ does not match the head of any clause in $K$, then the result of the query is undefined; alternatively, for an application, this can be set to $0.5$ (i.e., a random guess). 


\subsection{Uncertainty of a Knowledge Base}

Let $\mathcal{H}_K$ denote the set of the terms which are the heads of all clauses in $K$. We consider the heads of the clauses as these would correspond to the set of different queries that the knowledge base can compute a meaningful probability for.

Given a knowledge base $K$, we can then define an uncertainty measure $U_{KB}$ of $K$ as follows which takes into  account the entropy of answers it computes, i.e.,  the average  entropy of queries computable from $K$:
\be \label{Eq:AvgEntropy}
 U_{KB}(K) = \frac{1}{|\mathcal{H}_K|} \sum_{q \in \mathcal{H}_K} \sH_f^K(q).
\ee 
Ideally, if a knowledge base can answer all its queries with certainty (probability 1, i.e., true with probability 1 or false with probability 1), then $ U_{KB}(K)=0$ (assuming that $0\cdot \log(1/0)=0$), while it is $1$ in the worse case.
 
\begin{myexample}
Suppose we have a knowledge base $K$ as follows, in ProbLog:
\begin{verbatim}
0.2::a.
0.3::b.
0.5::a :- b.
\end{verbatim}
The set of the heads of all clauses in $K$ is $\{a,b\}$;  the possible queries $K$ can answer are $a$ and $b$, i.e.
$p[K \vdash a]= 1-(1-0.2)(1-(0.3)(0.5)) = 0.32$, and 
$p[K \vdash b]=0.3$. Thus,
\begin{align*}
 U_{KB}(K)  = \frac{1}{2} \left( \sH_f^K(a) + \sH_f^K(b) \right)  \approx 0.893
\end{align*}
\end{myexample}

\subsection{Sender's Message Choice Problem} \label{Sec:SenderChoice}

Communication plays a crucial role in reducing the uncertainty of a knowledge base. To illustrate this, suppose that Alice has a set $L$ of clauses and Bob has a knowledge base $K$. In order to minimize the average entropy of $K$ (with $m$), Alice can choose and send a message among those in $L$ to Bob, i.e.,
\be
m^* = \argmin_{m \in L} U_{KB}(K\cup\{m\}). \label{Eq:SenderChoice}
\ee

However, this requires Alice to have complete knowledge of~$K$. Alternatively, Alice might have a statistical approximation $A_i$ of $K$ in which $A_i \approx K$ with probability $p_{A_i}$. In this case, Alice's choice of message is recast as:
\be
m^* = \argmin_{m \in L} \sum_{i} p_{A_i} U_{KB}(A_i\cup\{m\}).
\ee
One way to realize this idea is allowing Bob to keep feeding the entropy of $K$ back to Alice. Then, throughout iterative communication, Alice can gradually improve $A_i$'s accuracy.







\subsection{Semantic Content of a Message} \label{Sec:EntropyChange}
We can define the notion of the {\em semantic content} $\mathcal{S}$ of  a message (where a message in this case is a clause labelled with a probability) with respect to the receiver's background knowledge base $K$ as follows, as the change in average entropy of a knowledge base with respect to its queries:
\be 
\mathcal{S}_K(m) =  U_{KB}(K\cup \{m\}) - U_{KB}(K)  .
    \label{EQ:S_K} 
\ee


Each message changes $U$ and the receiver wants 
to decrease the entropy, i.e., $\mathcal{S}_K(m) \leq 0$, or wants $\mathcal{S}_K(m)$ to be as low as possible, as the message $m$ should  decrease the average entropy in computed queries (of course, it could also increase the average entropy!).
In the following example, we show why this definition helps.

\begin{myexample} 
Suppose Alice has  a knowledge base $K$ as follows, in ProbLog:
\begin{verbatim}
0.3::b.
0.5::a :- b.
\end{verbatim} 
Suppose Alice receive the labelled clause $0.2::m$, i.e., $m$ labelled with probability $0.2$ forming $K'$ as follows:
\begin{verbatim}
0.3::b.
0.5::a :- b.
0.2::m.
\end{verbatim} 
Then, $p[K' \vdash a]=0.15$, $p[K' \vdash b]=0.3$, and $p[K' \vdash m]=0.2$ and so $U_{KB}(K') = \frac{1}{3} (0.60984 + 0.88129 + 0.721928) = 0.738$. We have:
\begin{align*}
\mathcal{S}_K(m)  & =   U_{KB}(K \cup \{0.2::m\}) - U_{KB}(K)  \\
  & = 0.738 - 0.746 \approx -0.008.
\end{align*}
The uncertainty in the knowledge base with respect to the queries it can answer has decreased - which is what we expect when Alice receives a clause with a lower entropy relative to the existing  clauses in $K$. 
Also,  if instead Alice received {\tt 0.9::b}, then Alice's knowledge base becomes:
\begin{verbatim}
0.9::b.
0.5::a :- b.
\end{verbatim} 
And $p[K' \vdash a]=0.45$, $p[K' \vdash b]=0.9$, that is, we have:
\begin{align*}
\mathcal{S}_K(m)  & = U_{KB}(K\cup\{0.9::b\}) - U_{KB}(K) \\
  & = 0.731 - 0.746 \approx -0.015 .
\end{align*}
The uncertainty in the knowledge base with respect to the queries it can answer has decreased - which is what we expect when Alice receives a clause with a lower entropy replacing an  existing  clause in $K$.
If we use ``$\cup$'' to assimilate {\tt 0.9::b}, then we have:
\begin{verbatim}
0.9::b.
0.3::b.
0.5::a :- b.
\end{verbatim} 
where  $p[K' \vdash a]=0.465$, $p[K' \vdash b]=0.93$, and
\begin{align*}
\mathcal{S}_K(m)  & = U_{KB}(K\cup\{0.9::b\}) - U_{KB}(K) \\
  & = 0.681 - 0.746 \approx -0.065 
\end{align*}
which is also a decrease in average entropy.
\end{myexample}

\subsection{Inference Can Reduce the Need for Communication} 
    \label{SS:IRNC}

In general, suppose there is no background knowledge, i.e., $K =\emptyset$, and the uncertainty of a query $q$ is $\sH^{\emptyset}_f(q) = 1$, i.e. the truth or falsity of $q$ is merely a random guess. But with a knowledge base $K\not= \emptyset$, we expect to have:
$\sH_f^K(q) \leq \sH^{\emptyset}_f(q)$.
Furthermore, for two different knowledge bases $K$ and $K^\prime$, if
$$
\sH_f^{K} (q) \le \sH_f^{K^\prime} (q),
$$
then we say $K$ is less uncertain than $K^{\prime}$ with respect to $q$. For $K^\prime \subseteq K$, we can easily show that 
$\sH_f^{K} (q) \le \sH_f^{K^\prime} (q)$.
 
This can lead to a reduction in the need to obtain information about $q$ given that we can make inferences about $q$ with $K$, e.g., suppose $\sH_f^K(q) > 1 - \delta$, where $\delta > 0$, is good enough, then there is no need to receive further information about $q$. In fact, with respect to $q$, we want only to receive information to reduce the entropy for $q$, that is, we want only to receive message $m \notin K$ such that:
\[
\sH^{K \oplus \{m\}}_f(q) \leq \sH^{K}_f(q) .
\]
This can also be generalized if there is a set of available messages, say $L$, as follows:
\be 
m^* = \argmin_{m \in L} \sH^{K \oplus \{m\}}_f(q).
    \label{EQ:mmin}
\ee 
Here, $m^*$ is the best message among those in $L$ to reduce the entropy for $q$. This implies that one might want to consider the consequences  of receiving and assimilating a message (or from the sender side, the implications of sending a message) on the uncertainty of a knowledge base (whether it would increase or decrease the entropy with respect to $q$ or with respect to the overall uncertainty of a knowledge base as defined above). 

\subsection{Improved Security via Semantic Messages}

As we have seen, the semantic content of a message helps reduce the receiver's uncertainty about one or more queries.
We can then define a notion of {\em semantically secure messages}, in that, without the receiver's knowledge base, someone who has gotten hold of the message might not be able to use it to answer a query (or a set of queries).

For example, suppose Eve has knowledge base $K_{\rm E}$ and Alice sends a message $m$ to Bob, who has knowledge base $K_{\rm B}$. With respect to a query $q$, we can represent the fact that Eve has little use for the message provided as follows:
\be
\sH_f^{K_{\rm E}}(q) = \sH_f^{K_{\rm E} \cup \{m\}}(q) .
    \label{sec:ignorance} 
\ee
In other words, suppose $\sH_f^{K_{\rm E}}(q) = 1$, and  Eve managed to intercept the communication and gain the message $m$ (and forwards it to Bob pretending that nothing has happened as a man-in-the-middle attack), but combined with her knowledge base $K_{\rm E}$, Eve is still just as uncertain about  $q$ as before.

However, Bob who receives $m$, who has $K_{\rm B}$, finds the message meaningful, that is, with respect to $q$:
\be
\sH_f^{K_{\rm B} \cup \{m\}}(q) <\sH_f^{K_{\rm B}}(q).
\label{sec:useful} 
\ee
Hence, as long as Bob and Alice have an a priori shared context, as represented by knowledge base $K_{\rm B}$ that Bob has and Alice knows that Bob has $K_{\rm B}$, then, it might be possible for Alice to transmit $m$ so that Eve (who does not know $K_{\rm B}$), an eavesdropper, will not be able to make much use of it, with respect to some ``sought  after'' answer for $q$.

Note that one can see this as analogous to the typical security encryption  scenario:  $q$ is the plaintext  encoded as the ciphertext $m$ using some key $k$, then Bob who has knowledge of the key $k$ can decrypt $m$ to know $q$, but Eve, after getting hold of $m$,  does not have $k$ and cannot use it obtain $q$. But there are key differences. There could be multiple ways to infer $q$ with different sets of clauses.
$K_{\rm B}$ and $K_{\rm E}$ may have different clauses but both could allow some inferences about $q$. 
Alice needs to ensure that $K_{\rm E}$ is such that \eqref{sec:ignorance} and $K_{\rm B}$ is such that  \eqref{sec:useful} before sending $m$.


We can consider semantic information security based on the previous discussion. Conventional information security \cite{Bloch11} \cite{Csiszar11} is based on different channel reliability (e.g., the eavesdropper channel is a degraded channel in wiretap channel models). 
On the other hand, semantic information security is based on the different reliability of knowledge bases.

\section{Key Issues in Designing SC Systems}

In this section, we discuss several issues in designing SC systems, in relation to the interactions between TC and SC.

\subsection{A Structure of SC with TC} \label{Sec:structured_SC}

\begin{figure}[h]
\begin{center}
\includegraphics[width=\figwidth]{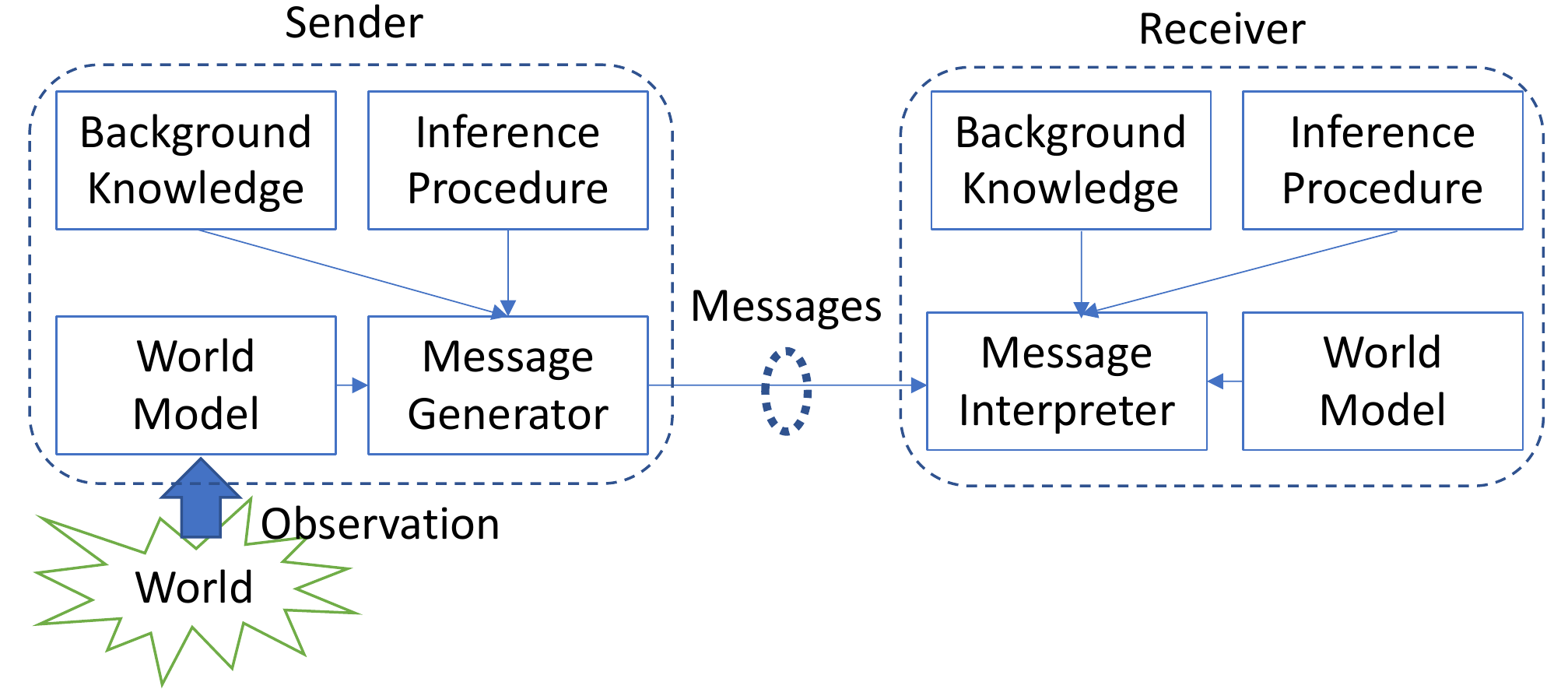} 
\end{center} \vspace{-20pt}
\caption{A model of SC from \cite{Bao11}.}
        \label{Fig:Fig1}
\end{figure}

In \cite{Bao11}, a model of SC  was presented, which is illustrated in Fig.~\ref{Fig:Fig1}. The message generator, which is also called a  semantic encoder is to produce a message syntax that will be transmitted by a conventional/technical transmitter. As a result, it is possible to design an SC system with two different layers: 
TC and SC layers.

\begin{figure}[h]
\begin{center}
\includegraphics[width=\figwidth]{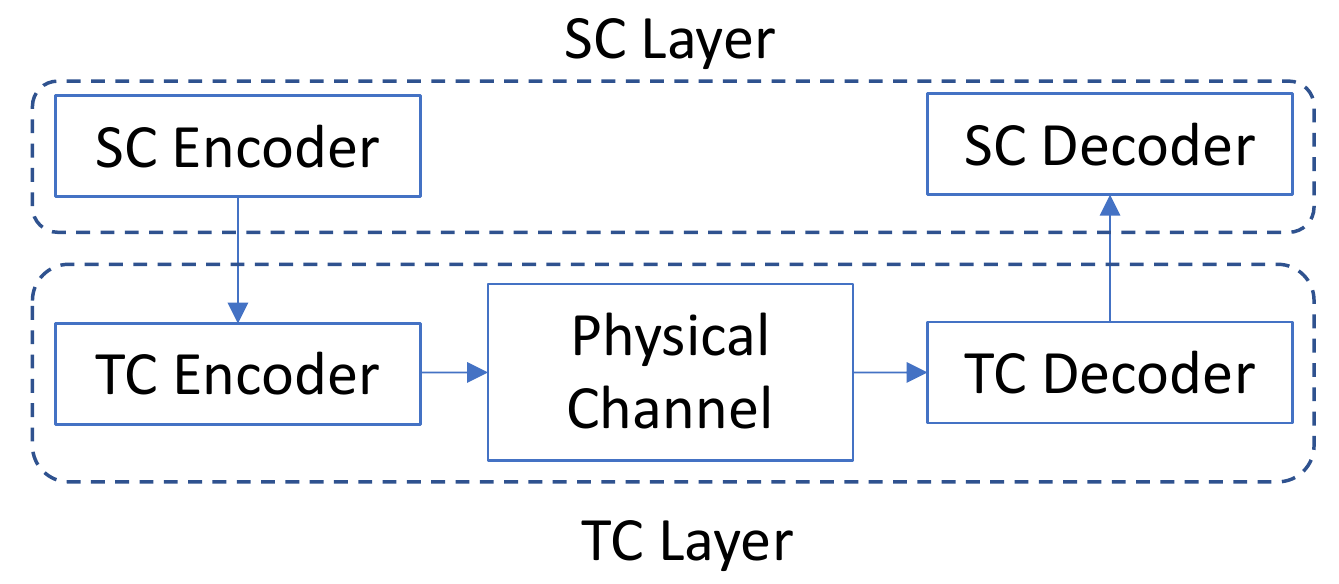} 
\end{center} \vspace{-10pt}
\caption{A two-layer model for SC over TC.}
        \label{Fig:Fig2}
\end{figure}

In particular, the output of the sender at the SC layer is a message to be transmitted over a conventional physical channel as shown in Fig.~\ref{Fig:Fig2}. The output of the decoder at the TC layer is a decoded message that becomes the input of the SC decoder. From this view, a conventional TC system can be used without any significant changes for SC. However, without any meaningful interactions between TC and SC, there is no way for TC to exploit the background knowledge in  SC and use the information obtained from semantic inference.

\begin{figure}[h]
\begin{center}
\includegraphics[width=\figwidth]{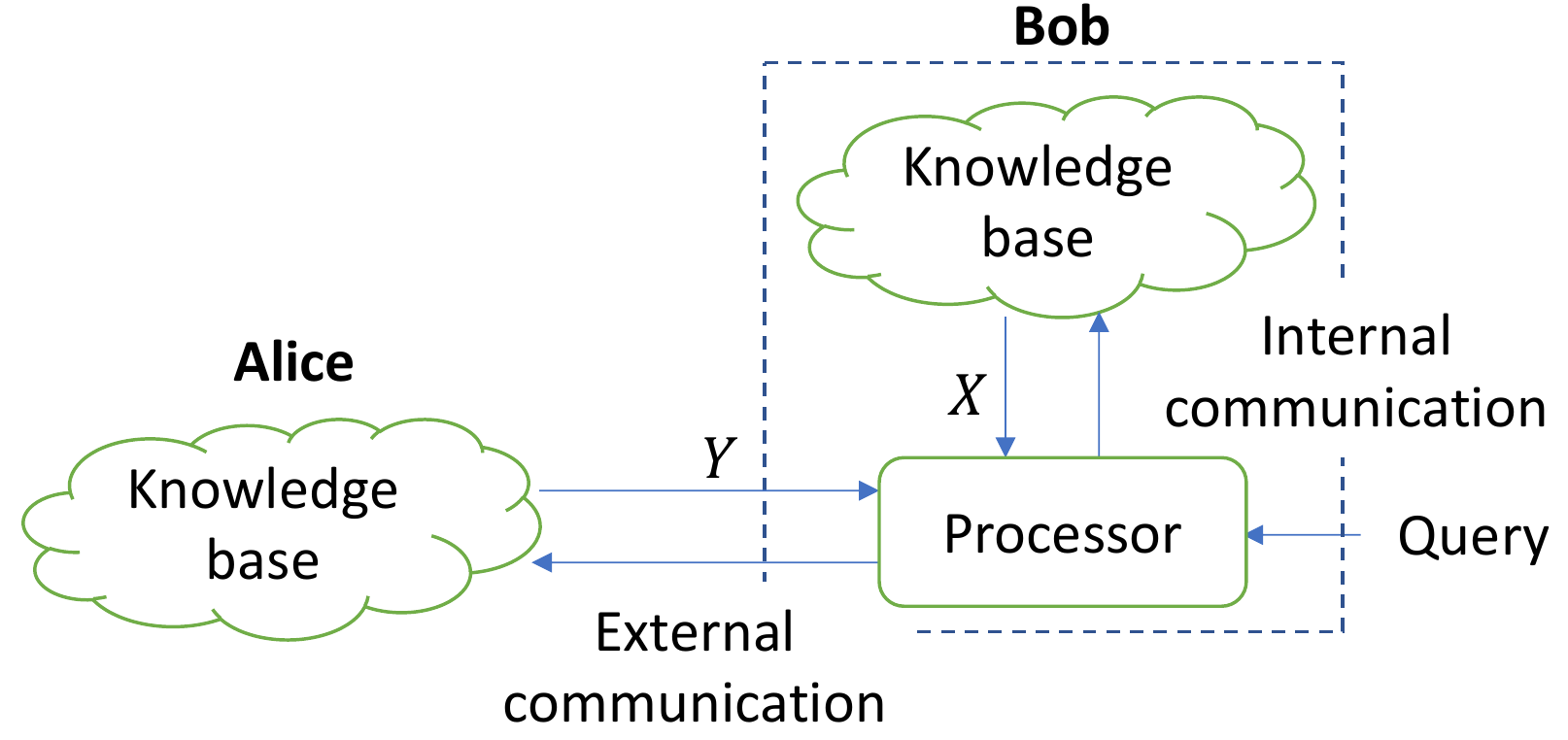} 
\end{center} \vspace{-10pt}
\caption{Exploiting the external and internal knowledge bases to reduce the number of bits to transmit.}
        \label{Fig:AB_XY}
\end{figure}

For interactions between TC and SC, the notion of the conditional entropy \cite{CoverBook} can be employed. 
In SC, we can assume that $X$ is the information that can be obtained from the background knowledge at the receiver. 
In particular, $X$ is a clause or an element of clauses in the knowledge base at the receiver. For a clause $X$, the entropy of $X$ becomes $\sH_f (X) = \sH(X)$.
In this case, the sender only needs to send the information of $Y$ at a rate of $\sH(Y\,|\,X)$. In Fig.~\ref{Fig:AB_XY}, we illustrate a model for exploiting the external and internal knowledge bases to reduce the number of bits to transmit. For a given query, Bob can extract partial information, $X$, from his knowledge base, which can be seen as data transmitted through internal communication, and seek additional information, $Y$, from others' knowledge bases, e.g., Alice's knowledge base. In this case, the number of bits to be transmitted is $\sH(Y\,|\, X)$, which will be available through external TC. 

\begin{myexample}
Suppose that Alice and Bob are the sender
and receiver, respectively. In previous conversations, Alice
told Bob that ``Tom has passed an exam and his score is 75
out of 100," which becomes part of background knowledge. Then, Bob asked Alice the pass score,  which is denoted by $Y$. Clearly, based on the knowledge base from the previous conversation, the pass score has to be less than or equal to 75, i.e., $Y \le 75$, which can be regarded as $X$. Thus, to encode $Y$, the number of bits becomes $\sH(Y\,|\, X)= \sH(Y\,|\, Y \le 75)$. If $Y$ is a positive integer and uniformly distributed over $[1, 100]$, $\sH(Y\,|\, Y \le 75) = 
\sum_{i=1}^{75} \frac{1}{75} \log 75 = \log 75$, not $\sH(Y) = \log 100$.
\end{myexample}

\begin{myexample}   \label{ex:str2}
Suppose that Eve told Bob that ``Tom's score is 75", which is denoted by fact $a$. In addition, Alice sends additional information that ``The pass score is 70," which is denoted by fact $b$.   Bob still does not know if Tom has passed, even after knowing the mark. Bob can ask Alice but does not need to ask Alice or Eve whether or not Tom has passed, because Bob can tell Tom passes from facts $a$ and $b$ via inference. If $p_a = 0.8$ and $p_b = 0.9$, the probability that Tom has passed is $p_a  p_b = 0.72$. Thus, in order to encode the fact that Tom has passed, which is a binary random variable 
(e.g., $Y = 0$ (resp. $Y = 1$) represents Tom passes 
(resp. fails)), the number of bits becomes 
$\sH_f (a \wedge b) = - (0.72 \log 0.72 + 0.28 \log 0.28) \approx 0.855 < 1$. This demonstrates that the background knowledge in SC can help compress the information in TC. 
A logic programming perspective on this example can also be considered. Suppose we model the knowledge Bob has with this rule that says that a person passes if the mark is above a threshold, and also that Bob has been told by Eve Tom's score:
{\small
\begin{verbatim}
0.8::mark(tom,75).
1.0::pass(X) :- mark(X,M), pass_score(S), M >=S.
\end{verbatim} 
}
But Bob still does not know if Tom has passed. Bob could ask Alice but does not need to if he also knows the passing mark:
{\small
\begin{verbatim}
0.9::pass_score(70).
0.8::mark(tom,75).
1.0::pass(X) :- mark(X,M), pass_score(S), M >=S.
\end{verbatim} 
}
 Bob can then answer the query {\tt pass(tom)} himself with computed probability $0.72$.
 Now Bob  knows not only Tom's mark but also whether Tom has passed, if this probability of $0.72$ is good enough for Bob. With $K$ representing Bob's knowledge base, note that $\sH^K_f({\tt pass(tom)})=0.593$. 
Note that if Charlie later tells Bob that Tom has passed with probability $0.6$, then Bob perhaps should discard Charlie's message which would increase Bob's uncertainty about {\tt pass(tom)} since $\sH^{K'}_f({\tt pass(tom)})=0.673$.
Inferring can go far - e.g., by inferring about Tom, Bob has reduced the need for communication, but this can be extended to not just Tom but many others, saving a lot of communication - another way to put it is that  suppose Bob knows the mark of 1000 students but without knowing the pass score, Bob does not know if any of them passed, but on receiving  the one message on the pass score, Bob now can infer which of the 1000 students passed and who did not. 
 Also, rather than sending $1000$ facts stating who passed and who didn't, sending just the pass score is more efficient. Lastly, if Bob is uncertainty tolerant and guesses the pass score $70$ with probability $0.75$, then it doesn't even need to ask for the pass score, and concludes Tom passes with probability $> 0.5$, which might be good enough for tolerant Bob.

\end{myexample}

In general,  the notion of the
Slepian-Wolf coding \cite{SW73} can be employed in order to efficiently exploit the background knowledge in SC. 
Suppose that there are two sources at two separate senders, which are denoted by $X$ and $Y$, for distributed source coding. In the Slepian-Wolf coding, sender 1 that has $X$ can transmit $X$ at a rate of $\sH (X)$, while sender 2 that has $Y$ can transmit $Y$ at a rate of $\sH(Y\,|\, X)$, not $\sH(Y)$. 
As a result, the total rate becomes 
$\sH(X) + \sH(Y\,|\, X) = \sH(X,Y) \le \sH(X) + \sH(Y)$. In the context of SC, $X$ can be seen as the information that is available from the background knowledge and through semantic inference. 

\subsection{SC for Efficient TC} 

As discussed in Subsection~\ref{SS:IRNC}, an optimal message can be chosen to minimize the entropy for a given query $q$  (see \eqref{EQ:mmin}). If a message is to be sent over a TC channel, the length of message can be regarded as the cost of TC. Let $L(m)$ denote the length of message for all available messages in $U$ at a sender (in bits),
while $K$ represents the knowledge base at the receiver that has query $q$.
Provided that the maximum length of message is limited by $L_{\rm max}$, the optimal message for query $q$ can be given by
\begin{eqnarray}
& m^* = \argmin_{m \in U} \sH_f^{K \cup \{m\}} (q) & \cr 
& \mbox{subject to} \ L(m) \le L_{\rm max}. &  
    \label{EQ:mLm}
\end{eqnarray}
While the optimization in \eqref{EQ:mLm} would be tractable, it requires for the sender to know or estimate the receiver's knowledge base, $K$, so that it can find $\sH_f^{K \cup \{m\}} (q)$. Thus, in general, it is expected that the sender has a larger knowledge base than the receiver and knows the receiver's knowledge base. For example, the sender can be a server in cloud and the receiver can be a mobile user in a cellular system. The server needs to update all the registered users' knowledge bases. In addition, the server is connected to base stations and needs to estimate the length of message $m$ to be transmitted through TC, which may vary depending on the time-varying physical channel condition between the user and associated base station. In this case, $L(m)$ is also a function of the channel condition and parameters of the physical layer (e.g., modulation order, code rate, and so on).

\subsection{Integration with Distributed Sources} \label{SS:DAS}

In this subsection, we discuss an approach to efficiently select distributed sources by minimizing the entropy gap, and extend it in the semantic context. 

Suppose that there are multiple senders and one receiver. Let $X_k$ denote the information that sender $k$ has. The receiver has a query and the answer is a function of the variables at the senders, which  is given by $
Y = \phi(X_1, \ldots, X_N)$,
where $N$ stands for the number of senders. 
For a large $N$, with a limited bandwidth, collecting all information from $N$ distributed senders may take a long time. Furthermore, if the $X_n$'s are correlated, it may not be necessary to collect all variables. For efficient data collection from distributed senders/sources (or sensors), 
the notion of data-aided sensing (DAS) has been considered in \cite{Choi_DAS19} \cite{Choi_DAS20}. If only one sender can be chosen in each round, the following selection criterion is proposed in \cite{Choi_WCNC20}:
\be 
n(i+1) = \argmin_{n \in \cI^c (i)} \sH(\cX^c(i)\,|\, \cX(i)) - \sH(X_n\,|\, \cX(i)),
    \label{EQ:ni}
\ee 
where $\cI (i)$ represents the index set of the senders that send their information up to iteration $i$ and $\cX(i)$ is the set of the variables of the senders corresponding to $\cI(i)$. Here, $\cX^c$ stands for the complement of a set $\cX$. In \eqref{EQ:ni}, $\sH(\cX^c(i)\,|\, \cX(i))$ represents the total amount of remained uncertainty of $\bx = [X_1, \ldots, X_N]$ for given 
$\cX(i)$, which is available at the receiver up to iteration $i$.  Thus, in the next iteration $i+1$, the sender that minimizes the remained uncertainty is to be chosen. 

While
no semantic information is taken into account in \eqref{EQ:ni}, it is possible to extend to consider semantic information. 
Let $m_n$ be the message at node $n$ (for a set of queries) and
$K(i)$ represent the updated knowledge base
at iteration $i$. Then, from \eqref{EQ:S_K}, the node (or source) selection criterion becomes:
\be 
n(i+1) = \argmin_{n \in \cI^c (i)} U_{KB} (K(i) \cup \{m_n\} ) - 
U_{KB} (K(i)) .
    \label{EQ:S_ni}
\ee 
That is, the receiver can actively seek the most effective message among multiple sources and iterate this process to rapidly improve the knowledge base. In addition, as in \eqref{EQ:mLm}, constraints on TC can be imposed if TC channels are limited (e.g., in terms of capacity and channel resource sharing).

\section{Conclusions and Open Issues} 

In this paper, we have proposed the SC layer that can simply be added on to the conventional TC layer. In addition, we have discussed how to jointly operate and interact such SC and TC layers with selected examples. While we have focused mainly on the Shannon-Weaver's semantics (Level B) problem, it has been presumed that all semantic contents can be useful for some generic tasks in terms of the effectiveness problem (Level C). However, such SC strategies may not be sustainable under limited memory for storing the ever-growing amount of knowledge, not to mention incurring redundant communication costs. To address this issue, an interesting topic for future research is to investigate the feedback and prediction mechanisms to estimate the semantic content’s \emph{task effectiveness} based on pragmatic information theory \cite{gernert2006pragmatic}, where we may first focus on a given task, and then count the usefulness of semantic contents based on its effectiveness in the task. In addition, there are recently proposed semantics-empowered and goal-oriented SC frameworks that commonly rest on AI-native operations with neural networks, as opposed to our knowledge-based SC layer. We expect that both AI-native and knowledge-based SC frameworks are complementary, even creating a  synergistic effect. This is another important open issue where effective integration in terms of various performance criteria are to be studied.

\bibliographystyle{ieeetr}
\bibliography{si}

\end{document}